# A Singular Perturbation Analysis for Unstable Systems with Convective Nonlinearity

Oliver Schönborn[a], Sanjay Puri[b] and Rashmi C. Desai[a]

[a] *Department of Physics, University of Toronto, Toronto, Ontario M5S 1A7 CANADA*
[b] *School of Physical Sciences, Jawaharlal Nehru University, New Delhi 110067, INDIA*

We use a singular perturbation method to study the interface dynamics of a non-conserved order parameter (NCOP) system, of the reaction-diffusion type, for the case where an external bias field or convection is present. We find that this method, developed by Kawasaki, Yalabik and Gunton for the time-dependant Ginzburg-Landau equation and used successfully on other NCOP systems, breaks down for our system when the strength of bias/convection gets large enough.

## I. INTRODUCTION

The study of non-equilibrium model systems has led to much progress in our understanding of self-organisation. This organisation emerges when the system is quenched to a region of its parameter space where competing mechanisms lead it to a new configuration, which may or may not be homogeneous in space or steady in time. The transition process often causes interfaces to appear between the allowed homogeneous states of the new configuration. These interfaces are diffuse on a molecular scale but appear as discontinuities on the mesoscopic length scale of the growing domains. The two disparate length scales make modelling of such systems and extraction of analytical results challenging. Here we will consider a model where mechanical transport competes with diffusive transport and dissipation.

An analytic method which approximately solves the nonlinear partial differential equations (PDEs) modelling some pattern forming systems, is a Singular Perturbation Method (SPM) developed by Kawasaki, Yalabik and Gunton (KYG) for the time-dependent Ginzburg-Landau equation [1]. It was subsequently applied to many other systems where the order parameter is non-conserved (NCOP) [1–3], conserved (COP) [4], coupled with long range repulsive interactions [5], *etc*, and which exhibit strong nonlinear behaviour at late times. However, the KYG method has not yet been applied, to our knowledge, to a system with a nonlinear convective derivative of the form $(\vec{v} \cdot \vec{\nabla})\vec{\psi}^n$, where $\vec{\psi}$ is the relevant order parameter field in the system. This type of nonlinearity occurs in hydrodynamics and also occurs when an external bias field $\vec{E}$ is present [6]. Then $\vec{E}$ plays the role of $\vec{v}$. Hence it is of some interest to examine the predictions of singular perturbation theory in this context.

To do this, we consider a simple model with a scalar NCOP – the Fisher Equation – and add a convective nonlinearity, whose strength can be tuned with a dimensionless parameter $\mu$. We call this new equation FEC (for Fisher Equation with Convective nonlinearity; *cf.* eq.(1)). The Fisher equation has been extensively studied in literature [7–9], originally in the context of population dynamics, and the KYG analysis yielded good results from its solution [7](*cf.* end of Introduction). For FEC with $\mu \ll 1$, the KYG SPM gives results very similar to those found for the Fisher equation, but breaks down at early times when $\mu \sim \mathcal{O}(1)$ or greater. This is because when $\mu$ is large enough, mechanical transport (which is a non-linear process) dominates over diffusive processes (which are linear). The KYG analysis can be expected to give usable results only when the front velocity is selected by the linear dynamics. This effect of the convective nonlinearity is expected to be general to other systems where bias field or convection is important.

Before we proceed to describe our results, some general remarks about the validity and utility of the KYG technique are in order. This technique is not meant to be a means of obtaining a solution to the initial-value problem for a reaction-diffusion equation. Rather, the singular perturbation method should be interpreted as a way of obtaining an analytic form, which may replicate the important features of the true solution, thereby enabling one to obtain statistically important quantities.

KYG [1] first applied the singular perturbation method to the Time-Dependent Ginzburg-Landau (TDGL) equation with a scalar order parameter. In this case, the analytic solution obtained is in disagreement on one important point with the real (obtained numerically) solution of the TDGL equation [7], *viz.* it has infinitesimally thin walls at late times, whereas the real solution always has walls of nonzero thickness. However, the analytic solution does reasonably reproduce the defect (interface) distribution of the real solution, starting from random initial conditions. Also the time-dependent structure factor calculated from the analytic solution of KYG [1] is identical to the better-known result of Ohta et al. [10], which is derived using interface dynamics. Thus, for the scalar TDGL equation, the singular perturbation approach is able to predict the "correct" asymptotic structure factor and the growth law for the characteristic domain size ($\sim t^{1/2}$, where $t$ is time). The next application of the SPM was to the $d$-dimensional Fisher equation [7,11]. As in the TDGL case, the interfacial profile of the KYG analytic solution (which is a travelling wave front) is artificially sharp compared to that for the real solution. However, the (travelling wave) analytic solution has the correct asymptotic velocity, even though the approach to this asymptotic velocity is incorrect in one dimension, where the exact result is known [9]. In the case where long-range interactions are also present, these statistics and defect dynamics are actually incorrectly given by the KYG method [12]. We find similar limitation for our *non-conserved* order-parameter system described by the FEC equation, in contrast to other non-conserved systems.

In Section II, we extend our earlier applications of the KYG method [2,5,7,11], to the FEC, and follow the formalism described in detail by Puri [7]. Because the KYG analysis can be applied to FEC with only small differences as compared to that done for the Fisher equation, only an outline of the method is given here, emphasising these differences, and the reader is refered to [7] and [1] for further details. In Section III, we compare the analytical solution with numerical results and summarize.

## II. MODEL: KYG METHOD ON FEC

The FEC in one space dimension is

$$\frac{\partial u}{\partial t} + \mu u \frac{\partial u}{\partial x} = \frac{\partial^2 u}{\partial x^2} + u - u^2, \qquad (1)$$

where the order parameter field $u$, the position $x$ and the time $t$ have been rescaled to dimensionless units. The dimensionless quantity $\mu$ is the ratio of convective to diffusive strengths in the equation of motion for $u$ and can not be scaled out. This equation is equivalent to Burgers equation [13] in which one would add a linear source term $u$ (providing the instability) and a quadratic sink term $u^2$ (providing the damping). Burgers equation is the simplest PDE to combine a nonlinear convective term with linear diffusion and is *analytically solvable*. It often appears as a limiting case in more complicated hydrodynamics problems, usually in the context of turbulence, and its main feature, due to the convective term, is that it yields shocks, also observed in the FEC. Note that $\mu = 0$ is the Fisher equation itself.

The essential idea of the KYG method is to generate an infinite order perturbative expansion around the linear solution of the PDE, approximate the $n^{th}$ order term in a suitable manner and resum the resulting infinite series to get an approximate, yet analytic, closed form result. There is an alternate derivation (also by KYG) using functional analysis that can be used for systems in which the non-linearities do not contain gradient terms. Since this is not the case here, we stick to the "standard" method.

The solution, in Fourier space, of the *linearized* equation (1) is $\tilde{u}_k^0(t) = e^{\gamma_k t} \tilde{u}_k(0)$ with $\gamma_k = 1 - k^2$. Thus for $k < 1$, $\tilde{u}_k^0(t)$ grows exponentially in time with rate $\gamma_k$. In what follows, the tilde denotes Fourier space and the superscript 0 denotes the solution to the linearized PDE.

The *non*linear equation (1) can be rewritten in an integral form in terms of the linear solution as

$$\tilde{u}_k(t) = \tilde{u}_k^0(t) - \frac{1}{(2\pi)^2} \int_0^t \int_{k_1,k_2} e^{\gamma_k(t-t')} \delta(k - k_1 - k_2) \times$$
$$g(k_1) \tilde{u}_{k_1}(t') \tilde{u}_{k_2}(t') dk_1 dk_2 dt'. \qquad (2)$$

where $g(k) = 1 - i\mu k$ and $\delta$ represents a Dirac delta function. The effect of convection is totally included in $g(k)$, which would be identically 1 for the Fisher equation. In applying the KYG technique, this modification creates no difficulty. Note however that if the nonlinear damping term had been a $u^3$ instead of a $u^2$, $g(k)$ would actually be a functional of $u$, so that the KYG analysis would be impossible to carry out.

We consider the integral in (2) as a singular perturbation: we introduce an expansion coefficient $\lambda$ in front of it and expand $\tilde{u}_k(t)$ in a power series in $\lambda$, $\tilde{u}_k(t) = \sum_{n=0}^{\infty} A_n(k,t) \lambda^n$. By equating the coefficients of equal powers of $\lambda$, expressions can be found for the $A_n(k,t)$ in terms of $\tilde{u}_k^0(t)$, and the result put in a diagrammatic form. Each $A_n$ grows exponentially faster in time than $A_{n-1}$. Higher order $A_n$ rapidly get hopelessly complicated so that each of the $n!$ diagrams making up a given $A_n(k,t)$ is approximated by the "comb" diagram (denoted $C_n^{\text{FEC}}(k,t)$) of the same order. The absolute error made in this approximation (which essentially gives equal weight to two- and multiple-$k$-mode time correlations) is uncontrolled and constitutes the most drastic approximation in the method.

At this point we have

$$A_n = n! C_n^{\text{FEC}}$$
$$= n!(-1)^n \prod_{i=1}^n \left\{ \int_{k_{2i-1},k_{2i}} \delta(k_{2i-2} - k_{2i-1} - k_{2i}) \times \right.$$
$$g(k_{2i-1}) \int_0^{t_{i-1}} dt_i e^{(t_{i-1}-t_i)\gamma_{k_{2i-2}}} \times$$
$$\left. \tilde{u}_{k_{2i-1}}^0(t_i) \tilde{u}_{k_{2i}}^0(t_i) \right\} \tilde{u}_{k_{2n}}^0(t_n). \qquad (3)$$

which is identical to the form obtained for the Fisher equation except for the factors $g(k_{2i-1})$ in the integral. Using a Laplace transform, (3) can be separated into $n+1$ terms, each growing exponentially in time. In order to simplify and to keep the expression tractable, a second approximation is made by KYG whereby only the fastest growing term of the sum is kept, which induces a *relative* error exponentially decreasing in time. This approximation is not affected by the form of $g(k)$ as long as $g(k)$ is independent of $\tilde{u}_k(t)$ (which is the case for FEC).

The integrand, now, is of the form

$$f(\{k_i\}) g(k_{2i-1}) e^{h(\{k_i\})} \tilde{u}_{k_{2i-1}}^0(0) \tilde{u}_{k_{2i}}^0(0),$$

where $f$ and $h$ are functions whose form are not important for now, and where $\{k_i\}$ denotes the set of $k_i$'s appearing in the integral. At late times one expects the maxima of $h$ to dominate the integrand, so one can apply the Laplace integral method to simplify it. One can show that the exponential is maximized when $k_{2i-1} = k/(n+1), \forall i$. There are several different ways of approximating the integrand, e.g. one can choose to evaluate only $f(\{k_i\})$ at the maximum, or both $f(\{k_i\})$ and $g(k_{2i-1})$, but only one yields a final result which is consistent with the initial PDE: one must evaluate both $f(\{k_i\})$ and $g(k_{2i-1})$ at the maximum of the exponential.

Applying these various steps to (3) yields:

$$C_n^{\text{FEC}} \simeq \left[g(\frac{k}{n+1})\right]^n \frac{(-1)^n}{n! \prod_{i=1}^n \left(1 + \frac{i+1}{(n+1)^2}k^2\right)} \times$$

$$\prod_{i=1}^n \left[\int_{k_i} \tilde{u}_{k_i}^0(t)\right] \tilde{u}_{k-\sum_1^n k_j}^0(t). \quad (4)$$

This means $C_n^{\text{FEC}} = \left[g(\frac{k}{n+1})\right]^n C_n^{\text{FE}}$, where the superscripts FEC and FE refer to the Fisher Equation with Convective nonlinearity and the Fisher Equation results for $C_n$, respectively. Even with the much simplified comb diagram (4), the power series for $\tilde{u}_k(t)$ is still not summable analytically. To do this and also be able to Fourier transform back to real space, one usually approximates the coefficient of the multiple integrals in (4),

$$\left[g(\frac{k}{n+1})\right]^n \frac{(-1)^n}{n! \prod_{i=1}^n \left(1 + \frac{i+1}{(n+1)^2}k^2\right)}, \quad (5)$$

to first order in $k$, since the major contribution to the Fourier transform comes from $k$ near 0 (for NCOP systems only). For $\mu = 0$ this is a very good approximation. But for $\mu \neq 0$, $g(k)$ is complex, so that a phase error is also introduced. Because our singular perturbation expansion contains an infinite number of terms which must partly cancel each other to give a convergent sum, this phase error, although small, could introduce cancellation (interference) effects whose consequences can be judged only *a posteriori*.

With these approximations, one can analytically resum the power series in $\lambda$ and invert the Fourier transform to get [15]

$$u^{\text{FEC}}(x,t) \simeq \left(1 + \mu\frac{\partial}{\partial x}\right) \left(\frac{u^0(x,t)}{1 + u^0(x,t)}\right). \quad (6)$$

Note that we recover the KYG solution to the Fisher equation when $\mu = 0$. The $\mu\frac{\partial}{\partial x}$ term induces the same type of asymmetry in (6) as that given to the PDE (1) by the convective term. However one of the consequences of $\mu\frac{\partial}{\partial x}$ is the development, at large times, of strong peaks in $u(x,t)$ at the interfaces, indicating a breakdown of the singular perturbation result. Furthermore, one can see that this occurs *for any finite order expansion* of (5) in $k$. Given that higher order diagrams ($n \to \infty$) will dominate the power series for $\tilde{u}_k(t)$ at large times, and that the $k$ values dominating the dynamics are near 0, a remedy would be to approximate $\left[g(\frac{k}{n+1})\right]^n$ with a simpler non-polynomial function in $k$, close to $\left[g(\frac{k}{n+1})\right]^n$ for $k \to 0$ and $n \to \infty$. The one possibility which will give tractable summation and a fourier-transformable diagram is $\exp(-i\mu k)$. This approximation, which is very good for $\mu \ll 1$ (*cf.* next section) exhibits no peaks at the interfaces even as $t \to \infty$. Then from (4) one now gets, instead of (6),

$$u^{\text{FEC}}(x,t) \simeq \exp(\mu\frac{\partial}{\partial x}) \left(\frac{u^0(x,t)}{1 + u^0(x,t)}\right). \quad (7)$$

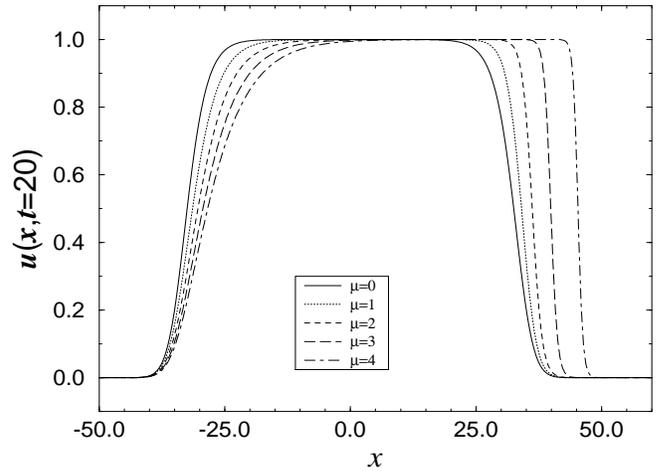

FIG. 1. Numerical solution of FEC at $t = 20$ and several values of $\mu$ (*cf.* figure legend), from a localised initial-condition in $x = 0$. The $\mu = 0$ curve is symmetric with respect to the origin. Others get distorted by the convection term ($\mu \neq 0$).

Although the $\exp(\mu\frac{\partial}{\partial x})$ operator does not create artificial peaks at the interfaces for any $\mu$ and $t$, it is in effect a translation operator. Hence if the KYG solution to the Fisher equation ($\mu = 0$) is denoted by $u^{\text{FE}}(x,t) = u^0(x,t)/(1+u^0(x,t))$, we have $u^{\text{FEC}}(x,t) = u^{\text{FE}}(x-\mu,t)$, i.e., the solution to the FEC we get from the singular perturbation expansion is in fact a translated solution to the Fisher equation. Since the true effect of the convective term can not be equated to a translation, it is immediately apparent that important characteristics of convection have been lost in this singular perturbation analysis. This will be crucial when $\mu$ is sufficiently large. We now briefly discuss and compare these analytic results with those of numerical integrations.

## III. DISCUSSION

In Fig. 1 we show the exact numerical integration of the FEC equation, at relatively small values of $\mu$ and intermediate times ($t = 20$), for a seed initial condition centered at $x = 0$ (i.e. $u(x, 0) = \delta(x)$). The $\mu = 0$ curve is a numerical solution to the Fisher equation. The standard Euler integration method was used for all numerical integrations, with time and space meshes suitably chosen to insure stability and precision. After a transient time approximately equal to 10, the interface profiles change little and the interface velocity reaches a value within 10% of its asymptotic value. Hence the curves for $t > 20$ are almost identical to those of Fig. 1 but with larger bulk region (where $u \sim 1$). As for the shock waves found in the solution of the Burgers equation [16], the effect of the convective term on the spatially symmetric solution of the Fisher equation is to steepen the right interface and broaden the left one, as $\mu$ increases. One way of seeing analytically how the convective nonlinearity operates is to look at the PDE $\frac{\partial}{\partial t} u = u \frac{\partial}{\partial x} u$ which has, in implicit form, $u(x, t) = f(x - u(x, t)t)$ as a solution. This is a "wave" with points at height $u$ travelling at speed $u$, so that higher points travel faster than lower ones.

In Fig. 2 we compare the results of a numerical integration for FEC with the analytical results of equations (6) and (7), for $\mu = 0.1$ and $t = 20$. For such a small $\mu$ the asymmetry is difficult to discern by eye for all three curves, but can be checked numerically. Also, because of the smallness of $\mu$, it is hard to distinguish between the solutions given by (6) and (7) (as expected from the approximations), so that one could use either (6) or (7) as approximate solutions to FEC, keeping in mind the remarks made at the end of last Section. The offset between the interface of either of the approximate solutions and that of the exact numerical one occurs because at early times the speed with which the fronts in the exact numerical solution travel is significantly smaller than that of the approximate solutions: the velocity of the fronts given by (6) and (7) are both equal to the *asymptotic* velocity of the exact solution, namely 2, as for the Fisher equation [7].

The interfaces given by our approximate solutions can be softened further by approximating $\prod_{i=1}^{n} \left(1 + \frac{i+1}{(n+1)^2} k^2\right)$ in (4) by $\exp(-k^2/2)$ instead of 1, which is an extremely good approximation (less than 1% error) for all $k \leq 1$ and $n \gtrsim 4$ (this softening is analogous to one done by Oono and Puri [17] and Puri [7]). However one finds that the change this induces in the profiles is very slight (not shown for that reason) and hence only useful at early times, so that the hardness of the interfaces of the approximate solutions is more deeply buried in the comb diagram approximation.

For $\mu \simeq 1$, the asymmetry between the two sides of the solution is much more obvious. But also (as shown in Fig. 3) eq. (7) gives a better solution than eq. (6). The steepness of the right interface of the eq. (6) solution at $t = 40$ indicates the beginning of the breakdown of the SPM result, since going to slightly larger times brings about a clear peak in the left interface and a dip in the right one. This does not happen for eq. (7). From the point of view of defect dynamics, the two solutions give the right velocity of 2 (obtainable from linear dynamics).

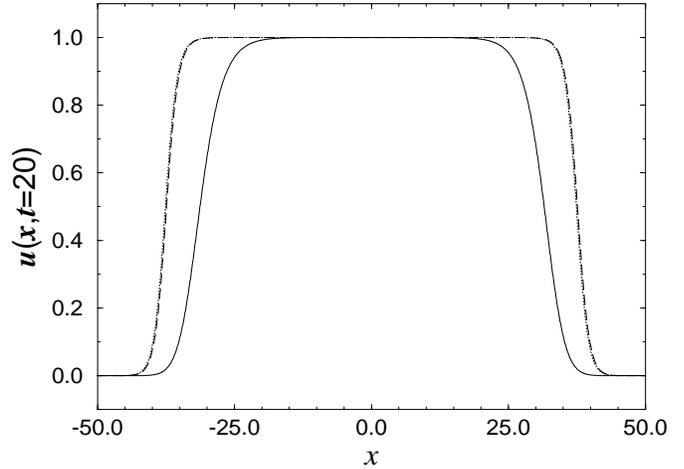

FIG. 2. Order parameter profile for the numerical solution (solid curve), for eq. (6) (dotted curve), and for eq. (7) (dashed curve), at $t = 20$ and $\mu = 0.1$.

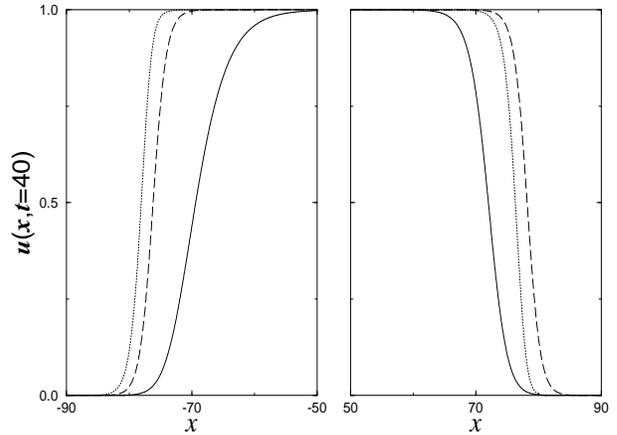

FIG. 3. Left and right interface profiles of the numerical solution (solid curve), of eq. (6) (dotted curve) and of eq. (7) (dashed curve), at $t = 40$ and $\mu = 1$.

For large $\mu$ (order of 10), both approximate solutions break down. The asymmetry of the analytical solution is clear but it matches in no way that of the real numerical solution. Indeed, the velocity of the right interface is wrongly given by both analytic solutions, as it is much greater than 2 (*cf.* Fig. 4) at all times. Peaks develop at the interfaces given by eq. (6) even at intermediate times like $t \simeq 20$ (not shown). Also, the left interface of the exact numerical result is extremely broad for $\mu$ on that order, something that the SPM seems incapable of describing. This can hardly be taken into account by the approximations used and, as can be seen, both eqs. (6) and (7) break down for the relative convective strength of the order of 10 (it actually breaks down for $\mu \simeq 5$ at $t \simeq 40$), not only in terms of order-parameter profile, but of defect dynamics.

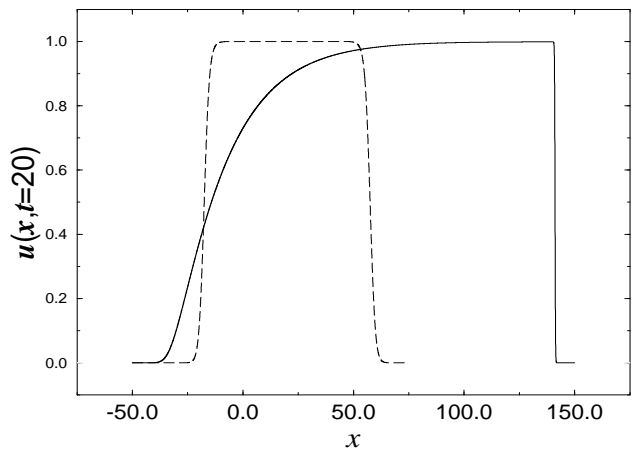

FIG. 4. Numerical solution (solid curve) and eq. (7) (dashed curve) at $t = 20$ and $\mu = 20$. Again, from initial condition in $x = 0$.

We can therefore say that KYG SPM gives results very close to the real solution in terms of profile and velocity up to $\mu \sim \mathcal{O}(1)$. But as the shock wave nature of the convective nonlinearity manifests itself more and more strongly (as $\mu$ increases), the KYG SPM does not manage to capture the essential features of convection, both in profiles and defect dynamics. Several analytical approaches are being investigated by us to make the results more quantitative and insightful. Apparently there is more to the failure of KYG at large $\mu$ than just a limitation of the SPM: any type of analysis performed around the *linear* solution, as done here, cannot be expected to work if the (attracting fixed point) solution becomes hyperbolic at some large enough value of $\mu$, while an alternate solution (corresponding to a different asymptotic profile and velocity) becomes stable.

## IV. CONCLUSION

The utility of singular perturbation methods lies in the calculation of statistical quantities (e.g., time-dependent structure factors, domain growth laws), which are determined by the qualitative features of the solution. However we have shown that the singular perturbation approach will not give a reasonable solution to the initial-value problem for a reaction-diffusion equation where convection is strong enough (a precise criterion will be given in a future article) even in terms of such statistical quantities since the defect dynamics are incorrectly predicted. It can be shown [18] that this incorrect defect dynamics is due to the convection term dominating the diffusive transport of perturbations, and is not due only to some intrinsic limitations of the singular perturbation approach itself. Hence approaches which treat the full non-linear PDE must be sought to study the dynamics of reaction-diffusion systems where an external bias field or convection is present.

## ACKNOWLEDGMENTS

The authors would like to thank Chuck Yeung for enlightening discussions. This work was partially supported by the Natural Sciences and Engineering Research Council and the Fonds Canadien pour l'Aide à la Recherche.